\documentclass[12pt,preprint]{aastex}

\shorttitle{Interaction of a Global EUV Wave and Coronal Structures}
\shortauthors{Li, et al.}

\begin{document}

\title{SDO/AIA Observations of Secondary Waves Generated by Interaction
of the 2011 June 7 Global EUV Wave With Solar Coronal Structures}

\author{Ting Li\altaffilmark{1}, Jun Zhang\altaffilmark{1}, Shuhong Yang\altaffilmark{1}, Wei Liu\altaffilmark{2,3}}

\altaffiltext{1}{Key Laboratory of Solar Activity, National
Astronomical Observatories, Chinese Academy of Sciences, Beijing
100012, China; [liting;zjun;shuhongyang]@nao.cas.cn}


\altaffiltext{2}{Lockheed Martin Solar and Astrophysics Laboratory,
Department ADBS, Building 252, 3251 Hanover Street, Palo Alto, CA
94304, USA}

\altaffiltext{3}{W. W. Hansen Experimental Physics Laboratory,
Stanford University, Stanford, CA 94305, USA}

\begin{abstract}

We present \emph{SDO}/AIA observations of the interaction of a
global EUV wave on 2011 June 7 with active regions (ARs), coronal
holes (CHs) and coronal bright structures. The primary global wave
has a three-dimensional dome shape, with propagation speeds ranging
from 430--780 km s$^{-1}$ in different directions. The primary
coronal wave runs in front of the expanding loops involved in the
CME and its propagation speeds are approximately constant within
10--20 minutes. Upon arrival at an AR on its path, the primary EUV
wave apparently disappears and a secondary wave rapidly reemerges 75
Mm within the AR boundary at a similar speed. When the EUV wave
encounters a coronal bright structure, an additional wave front
appears there and propagates in front of it at a velocity nearly a
factor of 2 faster. Reflected waves from a polar CH and a coronal
bright structure are observed and propagate fractionally slower than
the primary waves. Some of these phenomena can be equally explained
by either a wave or non-wave model alone. However, taken together,
these observations provide new evidence for the multitudes of global
EUV waves, in which a true MHD fast-mode wave or shock propagates in
front of an expanding CME bubble.

\end{abstract}


\section{Introduction}

Propagating global disturbances in the solar corona were first
detected by the \emph{Solar and Heliospheric Observatory}
(\emph{SOHO}) Extreme ultraviolet Imaging Telescope (EIT;
Delaboudini{\`e}re et al. 1995; Moses et al. 1997; Thompson et al.
1998) and have since then been commonly called ``EIT waves".
However, there has been no satisfactory explanation about the nature
of such disturbances. Many authors, such as Wang (2000), Warmuth et
al. (2001), and Patsourakos \& Vourlidas (2009), proposed that they
are truly fast-mode magnetohydrodynamic (MHD) waves, while others
(Delann\'{e}e 2000; Chen et al. 2002; Attrill et al. 2007a, 2007b;
Chen \& Wu 2011; Schrijver et al. 2011) suggested they are apparent
waves related to the opening and restructuring of magnetic field
lines caused by a coronal mass ejection (CME). A combination of both
true wave and non-wave mechanisms was then proposed to reconcile
these disparate models (Zhukov \& Auch\`{e}re 2004; Cohen et al.
2009; Liu et al. 2010; Downs et al. 2011). For recent reviews, we
refer to Vr\v{s}nak \& Cliver (2008), Wills-Davey \& Attrill (2009),
Gallagher \& Long (2011) and Zhukov (2011).

Global coronal waves have been observed to interact with various
coronal structures (Thompson et al. 1998, 1999; Delann\'{e}e \&
Aulanier 1999; Wills-Davey \& Thompson 1999; Veronig et al. 2006).
They tend to avoid strong magnetic fields in active regions (ARs),
stop at or partially penetrate into the boundaries of coronal holes
(CHs), and stop near a separatrix between ARs. Stopping at CHs was
confirmed by MHD simulations (Wang 2000; Wu et al. 2001). In
addition, Ofman \& Thompson (2002) found strong reflection and
refraction of the primary wave from an AR, as well as secondary
waves generated by the dynamic distortion of the AR magnetic field.
Due to EIT's low cadence (12$-$15 minutes) and TRACE's small field
of view, it was difficult to observationally test such numerical
simulations. This situation was partly alleviated with the launch of
the \emph{Solar-Terrestrial Relations Observatory} (\emph{STEREO};
Kaiser et al. 2008) Extreme Ultraviolet Imagers (EUVI; see Wuelser
et al. 2004). It detected the first example of reflection and
refraction of coronal waves at CH boundaries (Gopalswamy et al.
2009), which were recently numerically reproduced (Schmidt \& Ofman
2010). However, STEREO's improved, yet relatively low cadence of 2.5
minutes left room for potential ambiguity and led to debates about
the validity of this result (Attrill 2010).

The new Atmospheric Imaging Assembly (AIA; Lemen et al. 2011)
onboard the \emph{Solar Dynamics Observatory} (\emph{SDO}; Schwer et
al. 2002) takes full-disk images in 10 (E)UV channels at
1$\arcsec$.5 resolution and high cadence of 12 s. It thus offers
opportunities to elucidate previous ambiguities about the
interaction of a global EUV wave with ARs and CHs. We report here
such an example observed by AIA in unprecedented detail.

\section{Observations and Data Analysis}

On 2011 June 7, SDO/AIA observed a spectacular solar eruptive event
occurring in NOAA AR 11226 (S22W55). It involved an M2.5 flare, a
filament eruption, a CME and a global coronal EUV wave (Figure 1).
Among the 10 wavelengths of AIA, the 193, 171, 211 and 335 {\AA}
channels clearly show the global EUV wave that has a dome-like shape
(see also the wave on 2010 January 17 in Veronig et al. 2010) and we
focus on these channels in this study. The wave signature is similar
in the three hot channels, 193, 211 and 335 {\AA}, but different in
the cooler 171 {\AA} channel. The four channels correspond to
different temperatures: 193 {\AA} (Fe XII) at 1.5 MK (with a hot
contribution of Fe XXIV at 20 MK and cooler O V at 0.2 MK), 171
{\AA} (Fe IX) at 0.6 MK, 211 {\AA} (Fe XIV) at 2.0 MK and 335 {\AA}
(Fe XVI) at 2.5 MK (O'Dwyer et al. 2010; Boerner et al. 2011). The
coronal wave was also observed by the EUVI aboard \emph{STEREO A}
that was 95$\degr$ ahead of the Earth (Figure 6\emph{c}). EUVI
observes the chromosphere and corona in four spectral channels (304
{\AA}, 171 {\AA}, 195 {\AA} and 284 {\AA}) out to 1.7 $R_{sun}$ with
pixel size of 1$\arcsec$.6.

Here, we use a semi-automatic method (see Podladchikova \& Berghmans
2005; Liu et al. 2010) to obtain stack plots from 10$\degr$ wide
sectors on the solar surface (``A" and ``C"--``G", Figure 2). Each
data point in stack plots is the average of the pixels within a
sector along a circular arc at the same distance from the eruption
center (S22W55 for Sectors ``A", ``F" and ``G") or the initiation
site of secondary waves (N8W50 for Sectors ``C"--``E").

\section{Results}

\subsection{Erupting Loops and Primary Global Coronal EUV Wave}

At 193 {\AA}, a series of erupting loops started to expand at
06:20:00 UT (Figures 3\emph{b} and 5\emph{a}; Animation 2). They
appear as bright stripes with increasing slopes. A parabolic fit
indicates an acceleration of 86 m s$^{-2}$ and a final velocity of
280 km s$^{-1}$.

At 06:24:33 UT, the global coronal EUV wave initiated in front of
the erupting loops (Figures 1\emph{b} and \emph{c}; see the
Animations 1 and 2). Seen from the stack plots (Figures 3\emph{b}
and 5\emph{a}), the spatial separation between the coronal wave and
the expanding loops indicates that the former propagates in front of
the latter. In Sectors ``A" and ``F", the velocities of the wave are
respectively 510 and 430 km s$^{-1}$, and remain constant over the
propagation up to 350 Mm. The 211 and 335 {\AA} observations in
Sector ``A" are similar to those of 193 {\AA} (Figures 3\emph{d} and
\emph{e}). However, the wave appears different at cooler 171 {\AA}
(Figure 3\emph{c}). A group of small-scale coronal loops are
impacted by the coronal wave and expand with a speed of 130 km
s$^{-1}$. The expansion lasts about 3 minutes and then the loops
contract to their initial state. This is a possible indicator of
loop oscillations triggered by the passage of the EUV wave.

We also investigate the vertical propagation of the wave dome using
a straight slice (``B"; Figures 2 and 3\emph{g}). The initial
expansion of the loops is evident in the stack plot. Then it faded
away with height due to the steep radial intensity gradient. The
onset of the wave in this direction coincides with that of the
lateral wave. The upward expansion of the dome has a uniform
velocity of 730 km s$^{-1}$ projected onto the sky plane within the
AIA field of view, which is 1.5 times as fast as its lateral
expansion to the north, but comparable to that of its lateral
expansion toward the south. Note that the erupting filament material
is much slower with a large velocity range from 80 km s$^{-1}$ to
400 km s$^{-1}$, qualitatively similar to those of coronal jets
(e.g., Liu et al. 2009, their Figure 2(a)). The wave signal along
Slice ``B" is very weak at 171 {\AA} (Figure 3\emph{h}) and not
observed at 335 {\AA} (Figure 3\emph{j}).

\subsection{Secondary Wave Over AR 11228}

At 06:30:19 UT, the dome-shaped wave encountered a coronal bright
structure (denoted by ``1" in Figure 1\emph{a}) and the circular
wave front was deformed (Figure 1\emph{d}). Seen from the stack plot
of sector ``A" (Figures 3\emph{b}-\emph{e}), the intensity of the
coronal wave decreased rapidly after it passed through the coronal
bright structure, with an increased velocity of 1340 km s$^{-1}$
(Figure 3\emph{b}).

About 2 minutes later, the weak wave arrived at the boundary of AR
11228 and apparently stopped there (denoted by ``2" in Figure 2).
Almost simultaneously, a secondary wave appeared about 75 Mm into
the boundary, which is 506 Mm from the eruption center (Figure
1\emph{e} and Figures 3\emph{b}-\emph{e}, see the Animation 2). This
wave lasted 7 minutes and the velocity decreased from 500 km
s$^{-1}$ to 250 km s$^{-1}$ at 06:39:07 UT. The secondary wave is
different from the primary wave in shape and kinematics. The former
shows a clear deceleration, while the latter has a uniform velocity
during the majority of its lifetime. At 211 and 335 {\AA}, loop
brightening in AR 11228 is observed from 06:32 UT (Figures 3\emph{d}
and \emph{e}).

\subsection{Reflected Waves from Bright Structures and a Polar CH}

A reflected wave from the coronal bright structure ``1" (Figure
1\emph{a} and Figure 2, see the Animation 2) was observed between
06:33:46 UT and 06:42:10 UT (``R1" in Figures 1\emph{f} and
\emph{g}; Figures 3\emph{b}-\emph{e}). In order to investigate the
evolution of the reflected wave, we placed Sectors ``C"--``E"
starting at this structure (Figure 2) and obtained their running
difference stack plots as shown in Figure 4. Initially, the
reflected wavefront propagated at a speed of 140 km s$^{-1}$. At
06:36:10 UT, another wavefront was observed in front of it, and
propagated at a greater speed of 250 km s$^{-1}$ (Figure 1\emph{f}
and Figures 4\emph{e}-\emph{h}). The former and latter wavefronts
form a shape of ``bifurcation". About 3 minutes later, the former
wavefront faded below detection. The latter wavefront experienced an
evident deflection towards the east, and thus the later stage of the
wave can be seen in the stack plots of Sectors ``D" and ``E". Then
the wave disappeared at 06:42:10 UT.

A similar ``bifurcation" was also observed when the primary wave
approaches the coronal bright structure ``3" (Figures 1\emph{a} and
\emph{e}, see the Animation 2). As seen in the stack plot of Sector
``F" (Figures 5\emph{a}-\emph{d}), a new wavefront initiated in
front of it at 06:31:31 UT, and propagated at a higher speed
(\emph{v} $\sim$ 690 km s$^{-1}$). This new wave front was
subsequently deflected at the boundaries of ARs (see the red
segments in Figures 1\emph{d}, \emph{f} and \emph{g}), and its
propagating direction changed by 71$\degr$ in 9 minutes, similar to
what was observed by Attrill et al. (2007a) and was interpreted as
untwisting motions of a twisted flux rope involved in the CME. 171
observations show the wave front in Sector ``F" as darkening instead
of brightening (Figure 5\emph{b}), similar to that observed by
Wills-Davey \& Thompson (1999), Liu et al. (2010), Schrijver et al.
(2011) and Ma et al. (2011). This is suggestive of heating above the
characteristic temperature (0.6 MK) of the 171 {\AA} passband, as
simulated by Downs et al. (2011).

Another reflected wave from a southern polar CH (Figure 1\emph{a})
was also observed at 06:46:09 UT (Figures 1\emph{h} and \emph{i}).
Seen from the stack plot of Sector ``G" (Figures
5\emph{e}-\emph{h}), the primary wave propagated in the southward
direction with a speed of about 780 km s$^{-1}$. The reflected wave
from the boundary of the polar CH had a speed (\emph{v} $\sim$ 430
km s$^{-1}$) lower than that of the primary wave. At 06:51:07 UT,
the reflected wave was 320 Mm away from the eruption center, and the
velocity decreased to 110 km s$^{-1}$. The reflected wave from the
polar CH at 171 {\AA} appears as emission reduction (Figure
5\emph{f}).

\subsection{Three-dimensional Wave Dome}
Based on the observed evolution of the wave dome with \emph{SDO} and
\emph{STEREO A}, we constructed an approximate three-dimensional
shape of the wave dome. The dome is initially assumed to be a
spherical crown which is an opaque shell. Then we adjust the
parameters of the spherical crown (radius, height) until the
projections seen from the two viewpoints match the observed images.
The spherical crowns seen from the SDO and STEREO A viewpoints are
shown in right panels of Figure 6. The height of the spherical crown
is nearly 1.5 times as high as its radius (Figures 6\emph{b},
\emph{d} and \emph{f}). For example, at 06:25:30 UT, the radius of
the spherical crown was approximately 0.34 $R_{sun}$ and the height
was 0.51 $R_{sun}$. Its volume was about 3.3\% of the solar volume.
At 06:28:10 UT, the radius and height increased to 0.45 $R_{sun}$
and 0.675 $R_{sun}$, respectively, resulting in a volume increase of
nearly 2.4 times from 06:25:30 UT.

At 06:25:30 UT, the angle between the axis of the dome and the
radial direction of eruption center (S22W55) on the surface was
about 22$\degr$ (Figures 6\emph{b}, \emph{d}). About three minutes
later, the axis of the dome was inclined southward by 15$\degr$
(Figure 6\emph{f}), and approached the radial direction of eruption
center. In fact, the dome-shaped wave front propagates faster in the
vertical direction than the wave front on the solar surface (Figure
3; Veronig et al. 2010), and the spherical dome evolves gradually
into an ellipsoid. Full three-dimensional MHD modeling is required
to understand these observations.

\section{Summary and Discussion}

The SDO/AIA observations of the global coronal EUV wave on 2011 June
7 presented here have revealed several very interesting phenomena:

1. Upon arrival at an AR on its path, the primary EUV wave
apparently disappeared and a secondary wave rapidly reemerged 75 Mm
within the AR boundary with a similar speed of 500 km s$^{-1}$. We
speculate two parallel possibilities to explain this discontinuity
of wave propagation:

(1) The AR coronal loops can be perturbed by the impact of the wave
and then generate secondary waves, as found in MHD simulations
(Ofman \& Thompson 2002). In this case, the signal of impact is
expected to propagate as a fast-mode wave within the AR. The short
duration from the arrival of the primary wave at the AR boundary to
the onset of the secondary wave indicates a high velocity of 2400 km
s$^{-1}$, as marked by the green dashed line in Figure 3\emph{b},
which is in line with the AR fast-mode speeds found in observations
and simulations (Liu et al. 2011; Ofman \& Thompson 2002; Williams
et al. 2002). However, because the AR has a strong magnetic field
and high plasma density, the propagating fast-mode wave could
produce very little density and thus EUV intensity perturbation,
likely below detection such that the observed wave appears to vanish
in the AR.

(2) Another possibility involves the interpretation of the EUV wave
as a consequence of field line stretching (e.g., Chen et al. 2002,
their Figure 4). In this case, the AR lies below a separatrix
surface under overlying, larger-scale loops. When the expanding CME
pushes these overlying loops, the short AR loops underneath remain
unperturbed, leading to the apparent disappearance of the wave. When
the next long field line across the separatrix surface is pushed, a
new wave front reappears on the far side of the AR. Because of the
short distance between neighboring overlying field line across the
separatrix, this can result in a large apparent velocity of 2400 km
s$^{-1}$.

2. When the EUV wave encounters a bright coronal structure, i.e.,
small-scale, local loops (instead of large ARs), a new wave front
appears there and propagates in front of it at a velocity nearly a
factor of 2 faster (see Figures 4\emph{a}-\emph{h} and
5\emph{a}-\emph{d}). This is somewhat different from recent AIA
observations of fast wave fronts overtaking slow ones with a factor
of 2 difference in their speeds (Liu et al. 2010, their Figure 5),
but raises similar questions. If the observed wave fronts are indeed
MHD waves, their speeds would be determined by the medium in which
they propagate. Thus, the speeds along the same direction would be
identical, unless the leading wave has altered the local magnetic
field and plasma condition, say by compression (Harra et al. 2011),
and thus the characteristic wave speed. Our observations suggest
that the latter could be the case here. Another possibility is the
line of sight projection of waves at different heights. The faster
wave front at the bifurcation could result from an upward deflection
by the top portion of the small-scale loops toward greater heights,
while the slower wave front could be the refraction of the wave at
lower heights around these loops. The increase of Alfv\'{e}n speed
and fast-mode speed with height in the low corona within a few solar
radius can thus explain their different propagation speeds.
Alternatively, this could also be explained if these waves fronts
are associated with the expansion of the CME volume of multiple
layers at different speeds.

3. There are clearly signatures of reflected waves from a coronal
bright structure and a southern polar CH (Figures 3\emph{b}-\emph{e}
, 4 and 5\emph{e}-\emph{h}). Compared with earlier STEREO
observations of a similar event on 2007 May 19 (Gopalswamy et al.
2009), AIA's high cadence makes this a stronger case as
observational evidence that fast-mode (shock) waves are deflected
away from regions of high Alfv\'{e}n speed and reflected in regions
of large Alfv\'{e}n speed gradients (Uchida et al. 1973; Uchida
1974; Wu et al. 2001; Zhang et al. 2011). In recent MHD simulations,
Schmidt \& Ofman (2010) suggested that the reflected wave observed
by Gopalswamy et al. (2009) was triggered by the magnetic pressure
difference between the boundary of the CH and the primary wave, and
the excitation of secondary waves in different directions was caused
by the induced oscillation of the CH. In this event and the X2.2
flare on 2011 February 15, we find that the reflected waves always
propagate fractionally slower than the primary waves, while
Gopalswamy et al. (2009) found that the reflected wave in certain
directions are faster. In any case, this may suggest that the local
plasma condition and thus fast-mode wave speeds are changed upon the
passage of the primary wave, as we noted above.

4. We find that the wave has a three-dimensional dome shape, similar
to that reported by Veronig et al. (2010). The upward propagation
speed of the dome is 730 km s$^{-1}$, while its lateral expansion
speed ranges from 430 to 780 km s$^{-1}$, depending on the direction
(see also Ma et al. 2009). These speeds, a factor of 2 greater than
typical EUV wave speeds of 200--400 km s$^{-1}$ (Klassen et al.
2000; Thompson \& Myers 2009), fall in the range of the nonlinear
MHD wave/shock category proposed by Warmuth \& Mann (2011). In
addition, the wave front on the solar surface propagates at nearly
constant speeds in front of the expanding loops involved in the CME
that accelerates up to 280 km s$^{-1}$. This is consistent with
early \emph{STEREO} results (Kienreich et al. 2009 and Patsourakos
\& Vourlidas 2009), as confirmed with numerical simulations (Cohen
et al. 2009). Therefore, our observations suggest that the dome
could possibly represent a CME driven shock fronts, as recently
images by AIA (Ma et al. 2011).

In summary, we find a variety of phenomena associated with the
interaction of the global EUV wave with local coronal structures,
including secondary waves from ARs and small coronal loops (bright
structures), as well as reflected waves from CHs and coronal loops.
Some of these phenomena can be equally explained by either a wave or
non-wave model alone. However, taken together, these observations
provide new evidence for the multitudes of global EUV waves, in
which a true MHD fast-mode wave/shock propagates in front of an
expanding CME bubble (e.g., Zhukov \& Auch\`{e}re 2004; Cohen et al.
2009; Liu et al. 2010; Downs et al. 2011).

\acknowledgments {We acknowledge the SECCHI and AIA for providing
data. This work is supported by the National Natural Science
Foundations of China (40890161, 11025315, 10921303 and 11003026),
the CAS Project KJCX2-YW-T04, the National Basic Research Program of
China under grant 2011CB811403, and the Young Researcher Grant of
National Astronomical Observatories, Chinese Academy of Sciences.
Wei Liu was supported by AIA contract NNG04EA00C.}

{}

\clearpage

\begin{figure}
\centering
\includegraphics
[bb=60 185 535 658,clip,angle=0,scale=1.0]{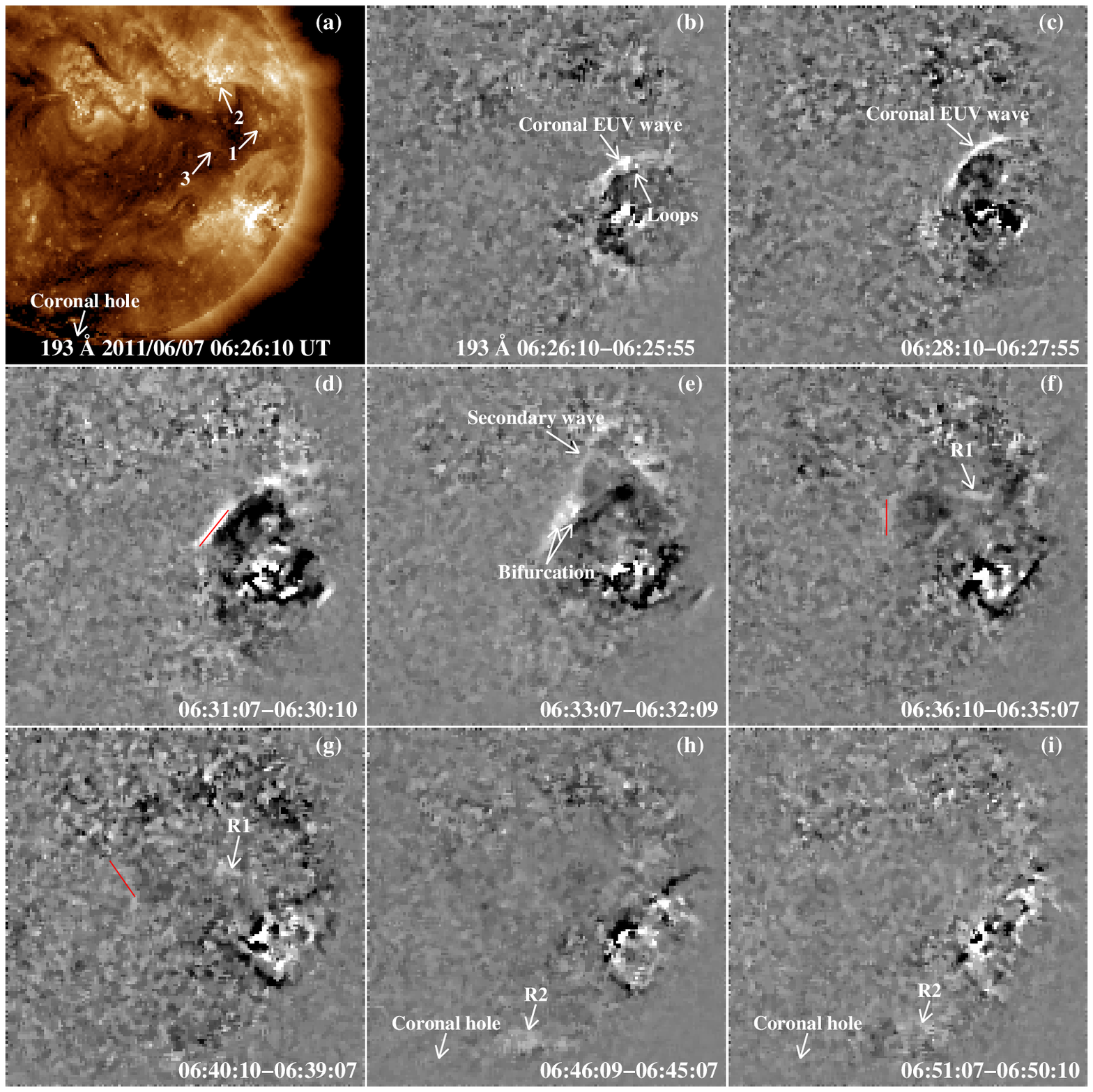} \caption{
{\emph{SDO}}$/$AIA 193 {\AA} image (panel \emph{a}, see the
Animation 1, available in the online edition of the journal) and
running difference images (panels \emph{b}--\emph{i}, see the
Animation 2) showing the evolution of the global coronal EUV wave on
2011 June 7. Arrows ``1" and ``3" in panel \emph{a} point to two
coronal bright structures hit by the coronal wave, and ``2" denotes
AR 11228. Red lines in panels \emph{d}, \emph{f} and \emph{g}
represent the leading edges, indicating the deflection of the wave.
``R1" and ``R2" denote the reflected waves from the coronal bright
structure ``1" and from the southern polar CH, respectively.
\label{fig1}}
\end{figure}
\clearpage

\begin{figure}
\centering
\includegraphics
[bb=74 198 522 640,clip,angle=0,scale=0.75]{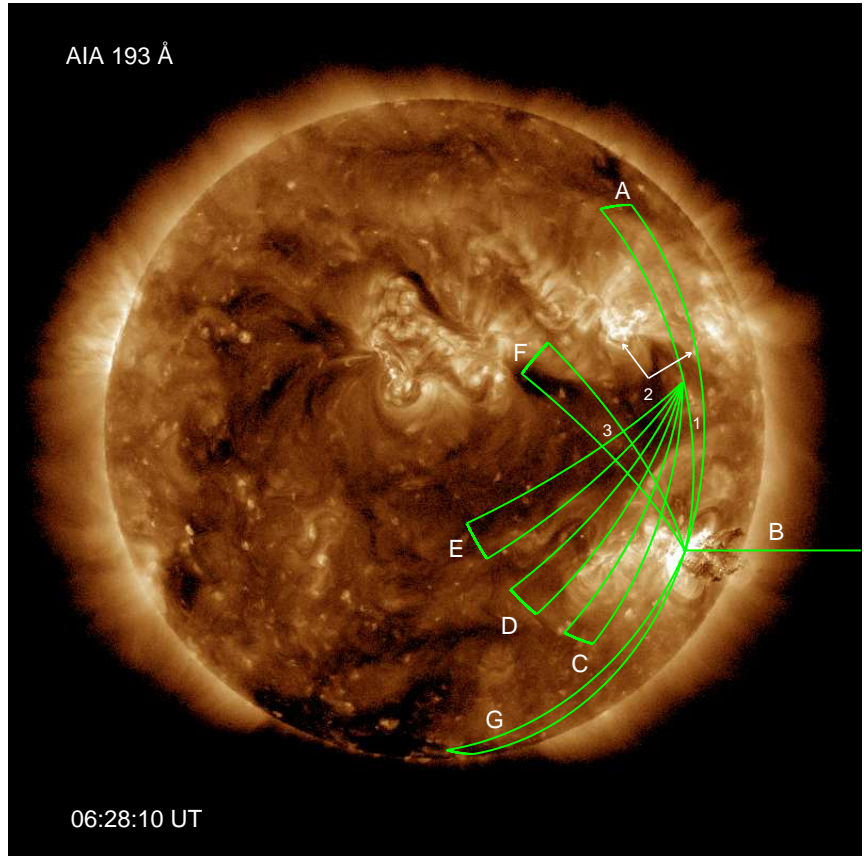}
\caption{{\emph{SDO}}$/$AIA 193 {\AA} full-disk image showing six
10$\degr$ wide sectors (``A" and ``C"--``G") and Slice ``B", which
are used to obtain the stack plots shown in Figures 3-5. ``1" and
``3" denote two coronal bright structures hit by the coronal wave,
and ``2" represents loop brightenings of AR 11228 (see Figures
3\emph{d} and \emph{e}). \label{fig2}}
\end{figure}
\clearpage

\begin{figure}
\centering
\includegraphics
[bb=43 80 522 740,clip,angle=0,scale=0.8]{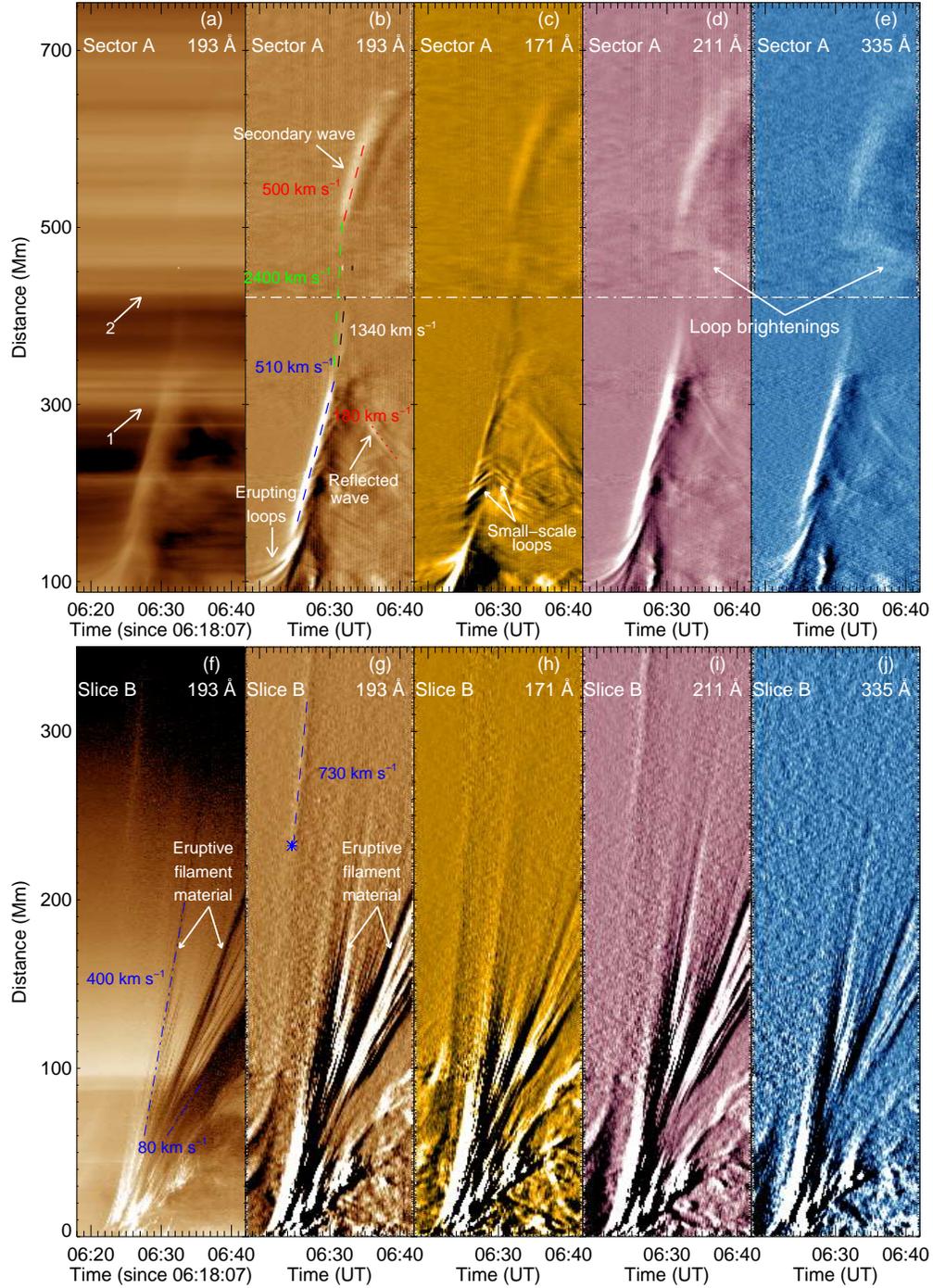} \caption{Original
and running difference stack plots along Sector ``A" and slice ``B"
at 193, 171, 211 and 335 {\AA}. ``1" denotes the coronal bright
structure hit by the coronal wave and ``2" represents AR 11228. In
order to visualize the structures better, the images above and below
the dash-dotted line are displayed with different color scales. The
asterisk in panel \emph{g} marks the start of the wave dome at the
limb. \label{fig3}}
\end{figure}
\clearpage

\begin{figure}
\centering
\includegraphics
[bb=99 100 476 718,clip,angle=0,scale=0.8]{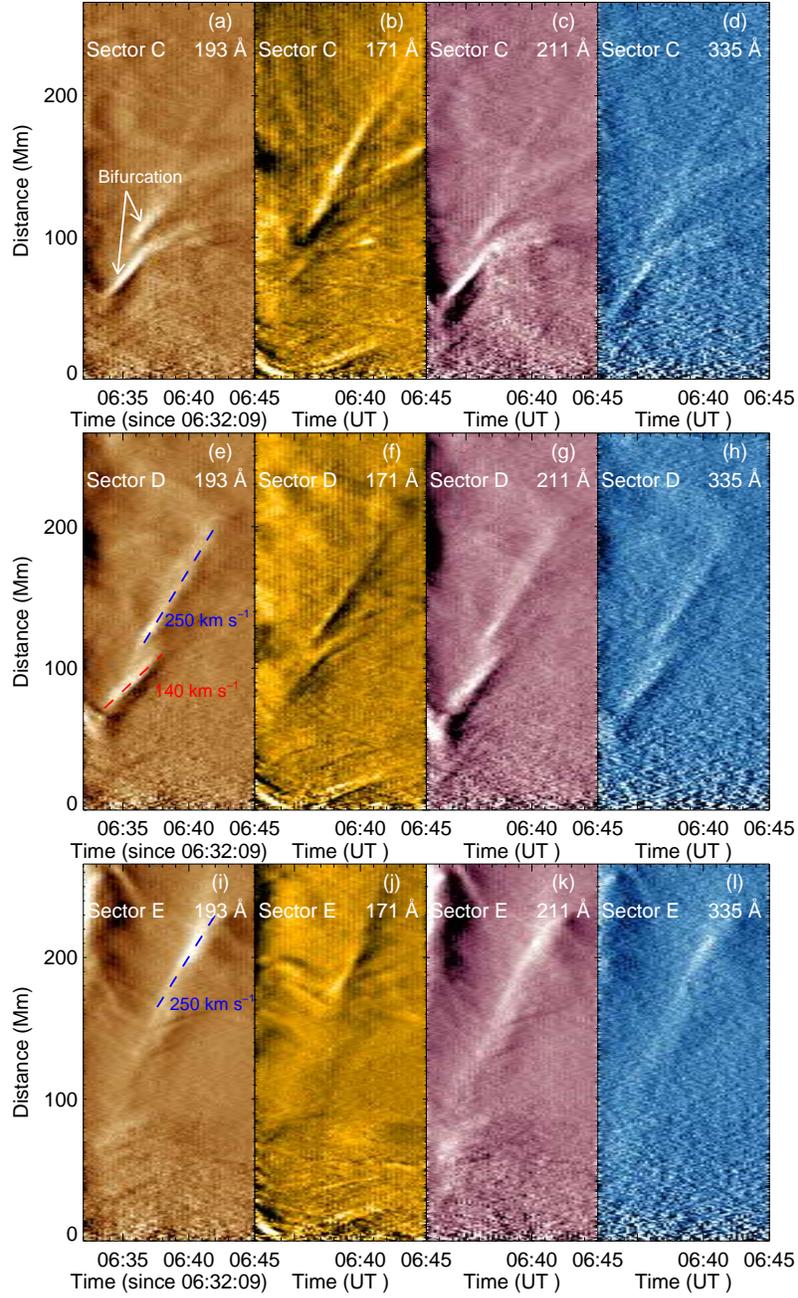} \caption{Running
difference stack plots at 193, 171, 211 and 335 {\AA} along Sector
``C" (top panels), ``D" (middle panels) and ``E" (bottom panels).
\label{fig4}}
\end{figure}
\clearpage

\begin{figure}
\centering
\includegraphics
[bb=102 142 477 678,clip,angle=0,scale=0.8]{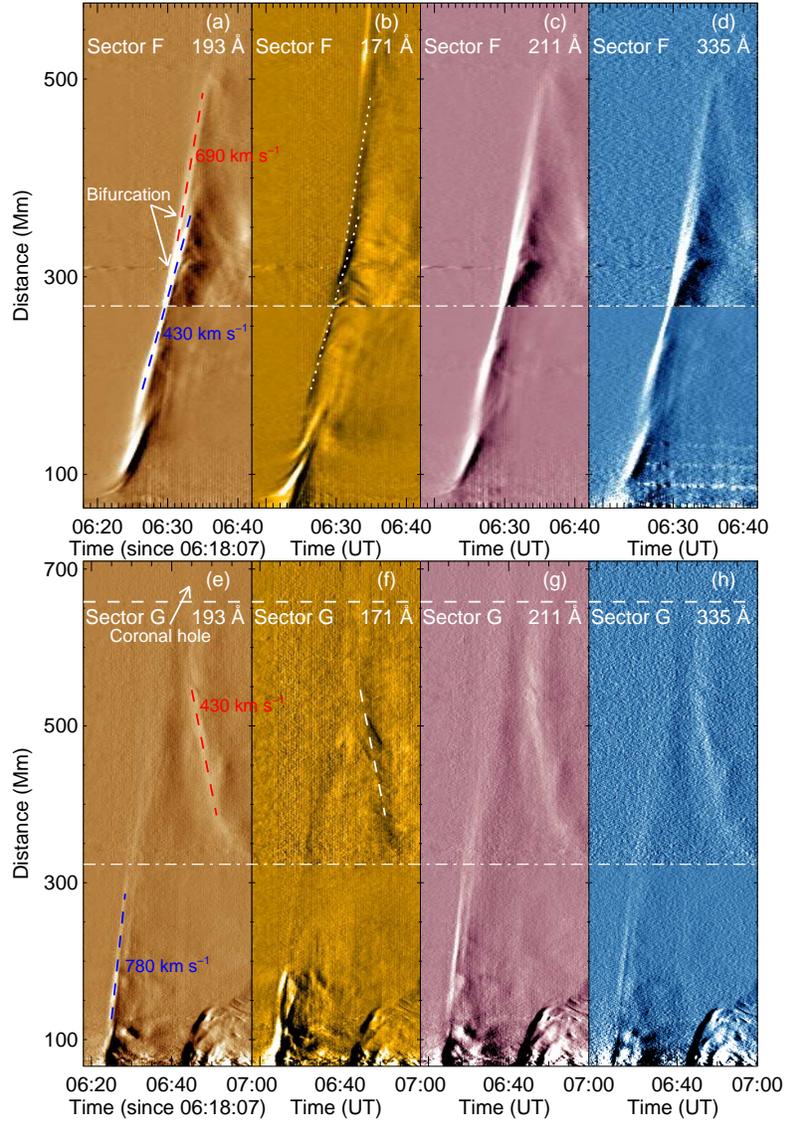} \caption{Running
difference stack plots at 193, 171, 211 and 335 {\AA} along Sector
``F" (top panels) and Sector ``G" (bottom panels). In order to
visualize the structures better, the images above and below the
dash-dotted line are displayed with different color scales. The
white dashed line in bottom panels represents the boundary of the
southern polar CH. \label{fig5}}
\end{figure}
\clearpage

\begin{figure}
\centering
\includegraphics
[bb=72 87 516 754,clip,angle=0,scale=0.8]{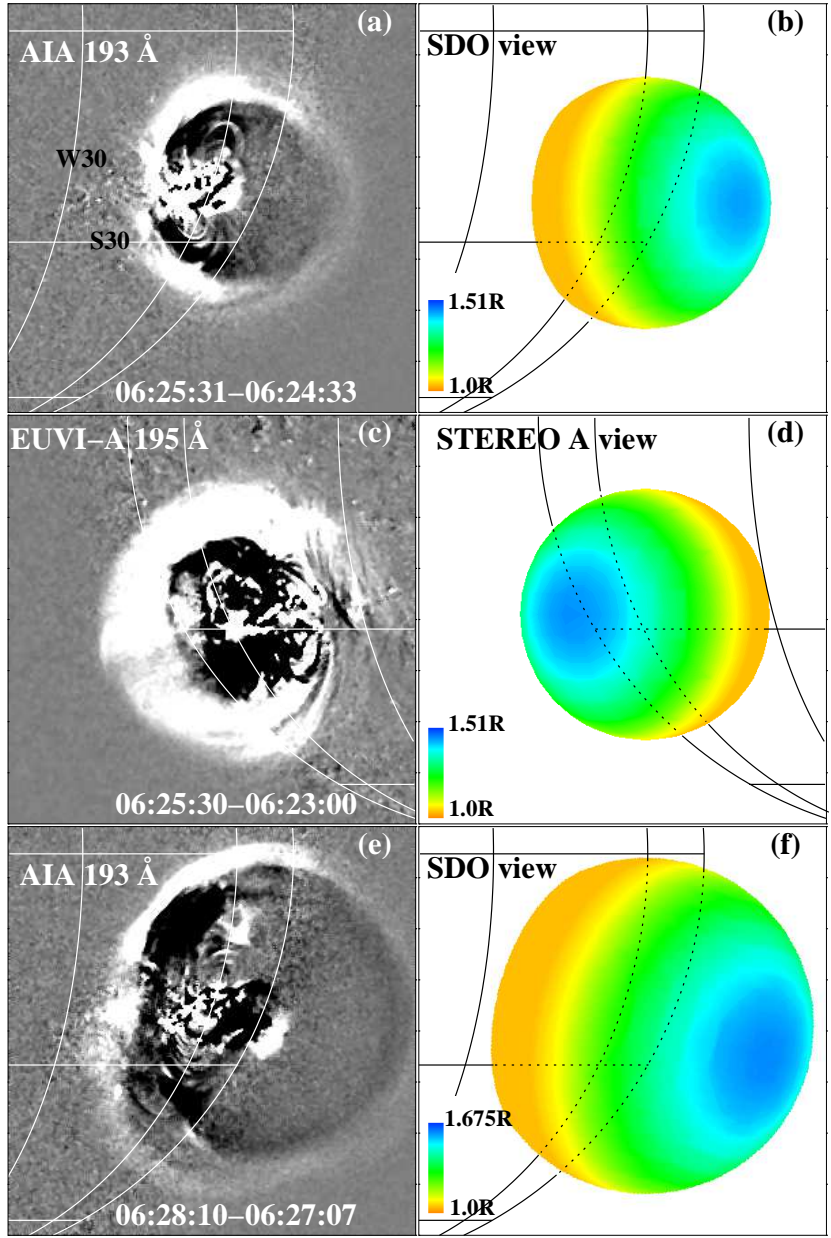} \caption{Observed
running difference images of AIA 193 {\AA} and EUVI-A 195 {\AA}
(left panels) and approximate three-dimensional shape of the wave
dome (right panels). Color represents height, and ``R" denotes the
solar radius. \label{fig6}}
\end{figure}
\clearpage


\end{document}